\documentclass[preprint,12pt]{elsarticle}

\usepackage{amssymb}

\journal{Journal of Magnetism and Magnetic Materials}

\begin{document}
\begin{frontmatter}



\title{Lifetime of skyrmions in discrete systems with infinitesimal lattice constant}


\author[inst1,inst2,inst3]{M.N.Potkina}

\affiliation[inst1]{organization={Department of Physics, St. Petersburg State University},
            city={St.Petersburg},
            postcode={198504}, 
            state={St.Petersburg},
            country={Russia}}
            
\affiliation[inst2]{organization={Department of Physics,  ITMO University},
            city={St.Petersburg},
            postcode={197101}, 
            country={Russia}}

\author[inst1,inst2]{I.S. Lobanov}

\affiliation[inst3]{organization={Science Institute and Faculty of Physical Sciences, University of Iceland},
            city={Reykjavik},
            postcode={VR-III, 107}, 
            country={Iceland}}

\author[inst3]{H. J\'onsson}

\author[inst1,inst2]{V.M. Uzdin}

\begin{abstract}
 Topological protection of chiral magnetic structures is investigated by taking a two-dimensional magnetic skyrmion as an example. The skyrmion lifetime is calculated based on harmonic transition state theory for a discrete lattice model using various values of the ratio of the lattice constant and the skyrmion size. Parameters of the system corresponding to exchange, anisotropy and Dzyaloshinsky-Moriya interaction are chosen in such a way as to keep the energy and size of the skyrmion unchanged for small values of the lattice constant, using scaling relations derived from continuous micromagnetic description. The number of magnetic moments included in the calculations reaches more than a million. The results indicate that in the limit of infinitesimal lattice constant, the energy barrier for skyrmion collapse approaches the Belavin-Polyakov lower bound of the energy of a topological soliton in the $\sigma$-model, the entropy contribution to the pre-exponential factor in the Arrhenius rate expression for collapse approaches a constant and the skyrmion lifetime can, for large enough number of spins, correspond to thermally stable skyrmion at room temperature even without magnetic dipole-dipole interaction.
\end{abstract}

\begin{graphicalabstract}
\includegraphics[width=0.8\textwidth]{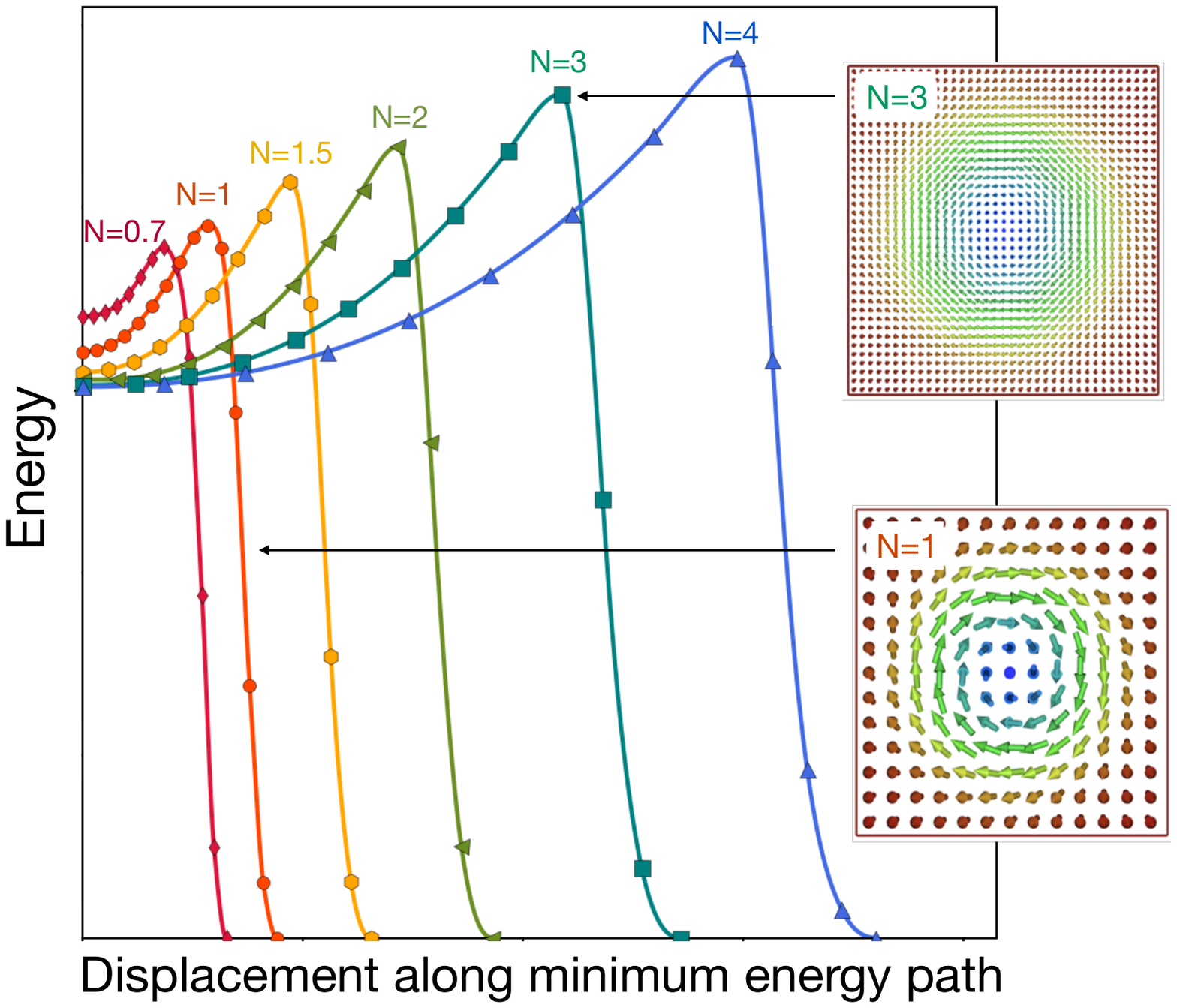}
\end{graphicalabstract}

\begin{highlights}

\item The lifetime of skyrmions is estimated within the framework of the harmonic transition-state theory
\item Collapse inside the sample and escape through the boundary are considered.
\item Pre-exponential factors and activation barriers are calculated for systems with different lattice constants corresponding to the same continuous micromagnetic model.
\item The manifestation of topological protection of chiral magnetic structures in discrete systems is shown.

\end{highlights}


\end{frontmatter}


\section{Introduction}
\label{sec:sample1}
Topological chiral magnetic structures, such as magnetic skyrmions, are promising information carriers in future generations of racetrack memory, logical and neuromorphic devices \cite{Finocchio16,Fert17,Everschor-Sitte18}. For such applications, thermal stability is a fundamental issue as thermal fluctuations and external disturbances can lead to structural breakdown and loss of the information encoded in the magnetic configurations. The theoretically predicted stability of topological magnetic systems is associated with the existence of a nonzero topological charge, an integer that cannot change under continuous variation of the magnetization. Magnetic skyrmion and homogeneous ferromagnetic states have different topological charges and, therefore, cannot be transformed into each other by continuous magnetization transformation \cite{Nagaosa13}. In real materials, magnetic moments are localized at discrete sites of the crystal lattice, and the topological arguments do, strictly speaking, not apply. Instead, one might expect to find finite energy barriers between states with different topological charges and if they are large enough the states can be protected against destruction by thermal fluctuations.
However, estimates of the energy barrier for the collapse of the small skyrmions in PdFe/Ir(111) \cite{Hagemeister15,Bessarab18,Uzdin18} and Co/Pt \cite{Lobanov16}, which have diameter 1-10 nm, are below 150 meV. This is of similar magnitude as energy barriers for magnetization reversal in small Fe islands on W(110) surface, where no  topological change is involved \cite{Krause09,Bessarab13}. The lifetime of nanoscale skyrmions studied so far
appears to be of the
same order of magnitude as that of non-topological magnetic structures of the same size. This means that despite the apparent topological protection, small skyrmions are quite unstable and can hardly exist at room temperature. 

Theoretical model calculations of skyrmion lifetime have, however, shown that it is in principle possible to identify material parameters giving stable skyrmions at room temperature. Parameter values can be found that lead to large enough energy barrier for skyrmion collapse and small enough pre-exponential factor in the Arrhenius expression for the lifetime  \cite{Varentcova20}. The question remains whether materials can be found that correspond to these parameter values. The manipulation of magnetic properties at atomic scale is a challenging task and the possible range for parameter values in quite narrow. On the micron scale for large skyrmions, effective parameters can be modified more easily. For example, in layered structures containing ferromagnetic (FM) layers and hard metal layers with large spin-orbit interaction, the effective Dzyaloshinsky-Moriya interaction (DMI) can be varied in a controlled way by the choice of materials and layer thickness. In such structures, skyrmion states have been found to be stable even above room temperature \cite{Woo16,Soumyanarayanan17}. Such modification of effective parameters that can lead to increased skyrmion stability is not directly related to topological protection. Moreover, as the number of magnetic moments in the skyrmion increases, the magnetic dipole interaction becomes more important and some have argued that it is the dipole interaction that is responsible for the observed stability of micron-scale skyrmions \cite{Buttner18}. Whether it is possible to have stability at room temperature in thin layers with negligible dipole interaction is still an open question. Furthermore, for skyrmions in antiferro- and ferrimagnets \cite{Potkina20}, as well as in synthetic antiferromagnets \cite{Legrand20}, where the magnetic dipole-dipole interaction is suppressed, the mechanism of 
stabilization still needs to be elucidated.

The stability of magnetic states depends not only on the activation energy of the transitions the system can undergo from these states but also on the pre-exponential factor in the Arrhenius expression for the rate of the magnetic transitions which depends on the
the shape of the energy surface in the vicinity of the initial and transition states \cite{Bessarab12,Coffey01}.
The role of the pre-exponential factor increases with increasing temperature. According to recent theoretical studies, \cite{Varentcova20} this can be used to create small 2D skyrmions that are stable at room temperature, although materials with the parameters proposed in \cite{Varentcova20}  have not yet been found. A change in the pre-exponential factor by more than 30 orders of magnitude for small changes in the magnetic field has been reported from experimental measurements \cite{Wild17}. 

When the size of a skyrmion is increased, the direction of neighboring magnetic moments becomes more similar and the continuous magnetization model becomes more appropriate for describing the system. How does topological protection present in continuous models manifest itself in the limit of infinitesimal lattice constant in discrete models? What limit does the energy barrier for skyrmion annihilation and the skyrmion lifetime approach when the ratio of the lattice constant and the skyrmion radius becomes infinitesimal? These questions are addressed in the present article.
We calculate the energy of the skyrmion state, the minimum energy path between the metastable skyrmion and the homogeneous FM state, and the activation energy of the skyrmion collapse as well as escape through a boundary for a gradually decreasing lattice constants while the skyrmion radius is kept constant. 
The pre-exponential factor in the Arrhenius for the life time is estimated within the harmonic approximation to transition state theory (HTST) \cite{Bessarab12} for several values of the lattice constant.  All parameters of the system, such as the exchange and anisotropy constants as well as the DMI are  chosen in such a way as to keep the size of the skyrmion and its energy unchanged for small enough lattice constant. The parameter values for the discrete lattice model are scaled with the lattice constant in such a way as to be consistent with the same continuous micromagnetic model. The calculated results can be interpreted either in terms of trends when the lattice constant is decreased and the skyrmion size remains the same, or in terms of increased skyrmion size if the lattice constant is taken to be the same. 

The article is organized as follows. First, in section 2. the model and the computational methodology is described. Then,
in section 3, the results are presented along with discussions.
Conclusions are given in section 4.


\section{Model and methods}

In this section the Hamiltonian of the systems studied is defined and the computational methods used to estimate the
energy barrier and pre-exponential factor in the Arrhenius expression for the life time described.


\subsection{Model}

The energy of the magnetic system is described by a Heisenberg-type expression for the energy of a two-dimensional (2D) square lattice

\begin{equation}\label{eq:heisenberg}
E[{\bf S}]=-\sum_{<j,k>}(J_{jk} {\bf S}_j\cdot {\bf S}_k+{\bf D}_{j,k}\cdot({\bf S}_j\times {\bf S}_k))
-K \sum_j {S^2_{j,z}}- \mu  \sum_j  {\bf B} \cdot {\bf S}_j,
\end{equation}
 where $ {\bf S}_i $ is a unit vector giving the direction of the magnetic moment at site {\it i}, $\mu$ is the value of magnetic moments taken to be the same on all sites, $J$ is the exchange parameter that is non-zero only for nearest-neighbour sites, and {\it K} is anisotropy parameter. The DMI  vector ${\bf D}_{j,k}$ is taken to lie in the plane of the lattice and point in the direction along the vector connecting atomic sites $j$ and $k$, thereby stabilizing a Bloch-type skyrmion. Formally, the lattice constant does not enter in eqn.~(\ref{eq:heisenberg}), but when the same system is described with higher resolution, i.e. smaller lattice constant, the parameter values corresponding to this system need to be changed in a certain way. In order to achieve this, a connection between the Heisenberg-type model 
 and a micromagnetic model with continuous distribution of magnetisation 
 is used. There, the energy with respect to the homogeneous FM state can be written as
\begin{equation}\label{eq:micromag}
E=\int \omega ({\bf r}) d^2{\bf r},  
\end{equation}
where
\begin{equation}
\omega ({\bf r})=  \mathcal{A} {\| \nabla \bf S(r) \|}^2 - {\mathcal{D}} ({\bf S(r)}\cdot \mathrm{rot}\bf S(r))
-\mathcal{K} ({\bf K^0}\cdot \bf S(r))- \mathcal{M} ({\bf B}\cdot \bf S(r))
\end{equation}
The integration is taken over the whole $\mathbb{R}^2$ space. The exchange stiffness $\mathcal{A}$, DMI density $\mathcal{D} $, anisotropy density $\mathcal{K}$ and magnetization $\mathcal{M}$ are proportional to the exchange parameter $\it J_{jk}$, modulus of the DMI vector ${\bf D}_{j,k}$, anisotropy constant $K$ and  $\mu$ in (\ref{eq:heisenberg}), respectively. The coefficients of proportionality depend on the type of crystal lattice, number of exchange parameters used in lattice model, {\it etc}. While the correspondence between continuous and lattice models is ambiguous in that different lattice parameters can correspond to a given micromagnetic model, the scaling of parameters
in the discrete model
is determined uniquely and depends only on the dimensionality of the system. For a 2D system, 
micromagnetic parameter values can be converted into parameter values for a discrete model of a square lattice with lattice constant $h$ by
%
%
$$J=2 \mathcal{A},\quad D=h \mathcal{D},\quad K =h^2 \mathcal{K},\quad \mu =h^2 \mathcal{M} .$$
This gives a scaling law for the values of the parameters as the lattice constant is changed in such a way that  correspondence is maintained to the same micromagnetic model
%
%
\begin{equation}\label{eq:scaling}
D\left(h/N\right)= \frac{D(h)}{N}, \quad K\left(h/N\right)= \frac{K(h)}{N^2},  \quad  \mu \left(h/N\right) = \frac{\mu(h)}{N^2} 
\end{equation}
while $J$ is independent of $N$.
Within this framework, calculations were performed for the following set of dimensionless parameters and $N=1$:
$D/J=0.35$, $K/J=0.16$, $\mu B/J= 0.02$. 
The cell size, $l=30$ ($30 \times 30$ spins subject to periodic boundary conditions) was chosen to be sufficiently large compared to the skyrmion size to prevent the influence of boundary conditions on the state of the skyrmion. In what follows, we will refer to $N$ as the scaling parameter, which denotes how the lattice constant is scaled.

In addition to the extended 2D system, calculations are also presented for a magnetic track with finite width in one dimension. Periodic boundary conditions are then only applied in the direction of the track while free boundaries are then used in the perpendicular direction, along the edges of the track. The escape of the skyrmion through the edge of the track is studied also as a function of $N$.


\subsection{Methodology}

The lifetime of the skyrmion state is estimated within the framework of HTST for magnetic degrees of freedom \cite{Bessarab12}.  Within this approximation, the lifetime, $\tau$, has an Arrhenius dependence on temperature 
\begin{equation}
\tau=\tau_0 e^{\Delta E / k_BT}
\label{arrhenius}
\end{equation}
where $\Delta E$ is the activation energy given by the energy difference between the saddle point and the initial state, $T$ is the absolute temperature and $\tau_0$ is the pre-exponential factor related to the entropy difference between the transition state and the initial state, as well as the flux through the transition state.
In a continuum description of the system, a smooth transformation of the magnetization connecting the skyrmion and FM states does not exist since these states have different topological charge \cite{Nagaosa13}. However, for a discrete lattice description, such a magnetic transition is possible for any lattice constant, but there can be a significant  energy barrier for annihilation making the skyrmion state stable for practical purposes. 
The question addressed here is how the activation energy and the pre-exponential factor depend on the scaling factor, especially in the limit of infinitesimal lattice constant which is the closest a discrete model comes to the continuum model.

To study this dependence, the minimum energy path (MEP) between the skyrmion and the FM state is calculated. The point of highest energy along the MEP, a first order saddle point on the multidimensional energy surface given by eq.~(\ref{eq:heisenberg}), 
minus the initial state energy gives the activation energy for the 
transition. For $N=1$ the dimensionality of the energy surface is already high, 1800, and it increases as the square of the scaling parameter, $N^2$.
For small $N$, the MEP can easily be calculated using the geodesic nudged elastic band method \cite{Bessarab15,Ivanov2020}.
When $N$ becomes large, this, however, becomes a computationally demanding task. Since the shape of the path is known it is possible to focus only on the region around the maximum and thereby reduce the effort significantly \cite{Lobanov17}. In this way calculations can be carried out for large values of $N$, even larger than $N=100$.

In HTST, the energy surface near the initial state and near the saddle point on the energy surface corresponding to the transition state
is approximated by a quadratic expansions. Zero modes corresponding to degrees of freedom for which the energy does not change significantly need to be treated separately. The expression for the attempt frequency $f_0=1 / \tau_0 $ is \cite{Bessarab12,Ivanov2017}
\begin{equation}
f_0= \frac{1}{2\pi}(2\pi k_B T)^{\frac{(P_{sk}-P_{sp})}{2}}\frac{V_{sp}}{V_{sk}} 
\sqrt{\sum\limits_{j=2}^{D-P_{sp}}{b_j^2}{\epsilon_{{sp},j}}}
\frac{\prod\limits_{j=1}^{D-P_{sk}}{\sqrt{\epsilon_{{sk},j}}}}{\prod\limits_{j=2}^{D-P{sp}}{\sqrt{\epsilon_{{sp},j}}}}
\label{eq:kHTST2}
\end{equation}
Here $\epsilon_{sk, j}$ and $\epsilon_{sp, j}$ are the eigenvalues of the Hessian matrix for the initial skyrmion state and at the saddle point, respectively, $V_{sk}$ and $ V_{sp}$ give the volume corresponding to zero modes while $ P_{sk}$ and $P_{sp}$ the number of such modes. The summation for the saddle point does not include the negative eigenvalue $\epsilon_{sp,1} $ and starts at $ j = 2 $.
The number $D$ in the upper limit of the summations is equal to the dimension of the energy surface $2N^2l^2$. The variables $b_j$ denote the expansion coefficients for the unstable mode at the saddle point derived from the linearized Landau-Lifshitz equation of motion.  More precisely $ b_j= \frac{\gamma}{\mu} 
\left({\bf e}_{sp,1} \cdot \left[ {\bf S}_{sp}\times {\bf e}_{sp,j} \right]\right)$,
where $ \gamma$ is the gyromagnetic ratio, $\mu$ the length of the magnetic moments, ${\bf e}_{sp,1}$ and $ {\bf e}_{sp,j}$ are the unit vectors along the first eigenvalue (normal to dividing surface) and the j-th eigenvector of the Hessian at the saddle point, and ${\bf S}_{sp}$ corresponds to the spin configuration at the saddle point \cite{Lobanov20}.

First, consider the entropic contribution to the expression for $f_0$ with the product of square roots of the positive eigenvalues of the Hessian in the numerator (for skyrmion state) and denominator (for transition state)(\ref{eq:kHTST2}). These expressions can be rewritten as
\begin{equation}
\prod\limits_{i}^{D^\ast} \sqrt{\epsilon_i}= \exp \left( \frac{1}{2} \sum\limits_{i}^{D^\ast} \ln{\epsilon_i} \right) = \left[ \exp \left( \frac{1}{D^\ast}\sum\limits_{i}^{D^\ast} \frac{1}{2}\ln{\epsilon_i}\right) \right]^{D^\ast}
\label{epsilon}
\end{equation}
The upper limit $D^\ast $ in eq. (\ref{epsilon}) takes into account the absence of zero modes at the saddle point and in the skyrmion state, as well as the negative eigenvalue at the saddle point.

Recently, another algorithm for calculating the entropic contribution to the pre-exponential factor has been presented \cite{Lobanov20}. This contribution can be written via determinants of the Hessian at the minimum and saddle point. An algorithm for calculating these determinants without searching for eigenvalues of the Hessian but using recursive relations makes use of the sparse structure of the Hessian matrix for energy in the Heisenberg-like energy expression when only short-range interactions are included.
The method makes it possible to calculate determinants for systems containing millions of magnetic moments. This makes it possible to calculate the pre-exponential factor and estimate the lifetimes of micron-scale topological structures with atomic resolution 
\cite{Potkina2020b}.


\section{Results and discussion}

Results are presented from calculations of both activation energy and pre-exponential factor in the Arrhenius expression for the life time of the skyrmion. Two annihilation mechanisms are considered,
radial collapse within an extended 2D system and escape through the edge of a track with finite width.

\subsection{Activation energy}

\begin{figure}[]
\centering
\includegraphics[width=0.8\textwidth]
{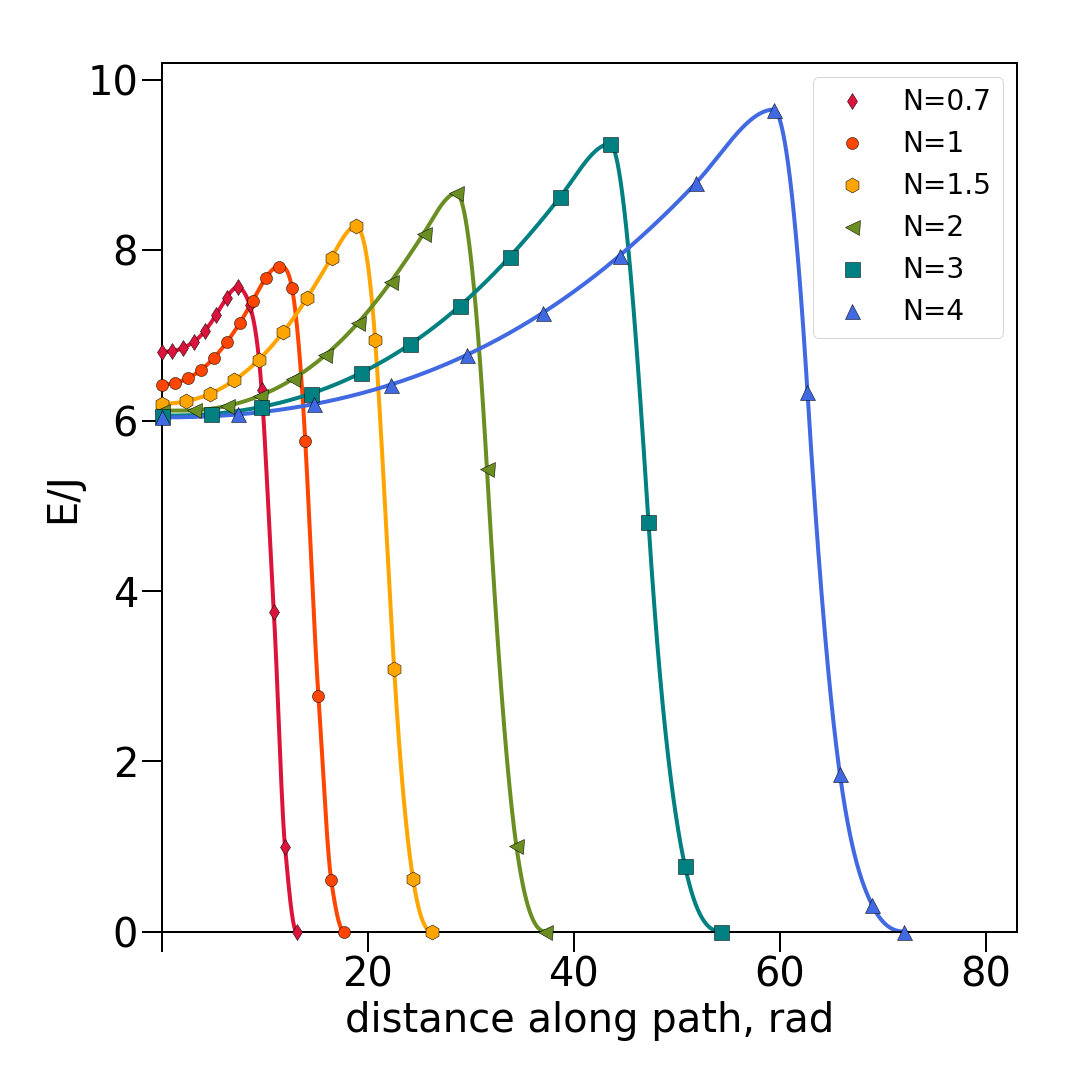}
\caption{Minimum energy path for radial skyrmion collapse calculated for several values of the scaling parameter.
The activation energy for collapse increases as the lattice constant is decreased over this range of small values of $N$ but reaches a large $N$ saturation limit at $N \approx 100$ where the energy maximum is $4\pi J$ with respect to the ferromagnetic state which is taken to give the zero of energy in each case. }
\label{MEP-collapse}
\end{figure}

Figure \ref{MEP-collapse} shows the computed MEPs for radial skyrmion collapse for several values of $N$. The path length represents the total change in orientation of all the spins and therefore increases with $N$ since a larger number of spins then forms the skyrmion structure. 
The energy of the initial skyrmion state first decreases with $N$, but already at $N > 2$ the change becomes insignificant.
The energy maximum along the MEP increases with scaling parameter and is still increasing for $N=4$.
But, calculations for much larger values of $N$ show that at $N \approx 100$ and larger the maximum energy changes insignificantly and approaches the lower bound of the energy of a topological soliton given by the $\sigma$-model \cite{Belavin75}.
This can be understood from the fact that, according to the scaling relations in eq. (\ref{eq:scaling}), all the parameters of the expression for the energy in eq. (\ref{eq:heisenberg}), with the exception of the exchange parameter $J$, tend to zero with increasing $N$.
Nevertheless, the contribution of various terms in eq. (\ref{eq:heisenberg}) to the energy of the skyrmion state does not change since the number of magnetic moments that form the skyrmion also increases with increasing $N$.
However, the number of lattice sites corresponding to noncollinear ordering at the saddle point is much smaller than that of the skyrmion and it increases relatively slowly with increasing $N$. Therefore, the transition state for large $N$ is a topological soliton that can be described within the $\sigma$-model \cite{Yang01} which only includes exchange interaction.
In the continuous $\sigma$-model the energy $E$ of any structure with topological charge $q$ satisfies the inequality $E\geqslant4\pi Jq$ \cite{Belavin75}. Consequently, the energy barrier for the nucleation of a structure with $q=1$ cannot be lower than $4\pi J$.
Our calculations show that the energy barrier is nearly equal to this value for large $N$ and that the barrier for the nucleation of a skyrmion from the FM state in the continuous case can be estimated from the soliton energy in the $\sigma$-model. The collapse of a skyrmion passes through a state corresponding to this solution with a small radius (infinitesimal in the continuous limit).


For a track of finite width, a skyrmion can also be eliminated by escape through the boundary \cite{Bessarab18,Uzdin18=PhB, Cortes17}. This mechanism corresponds to another path on the energy surface.
%
\begin{figure}[h]
\centering
\includegraphics[width=0.7\textwidth]
{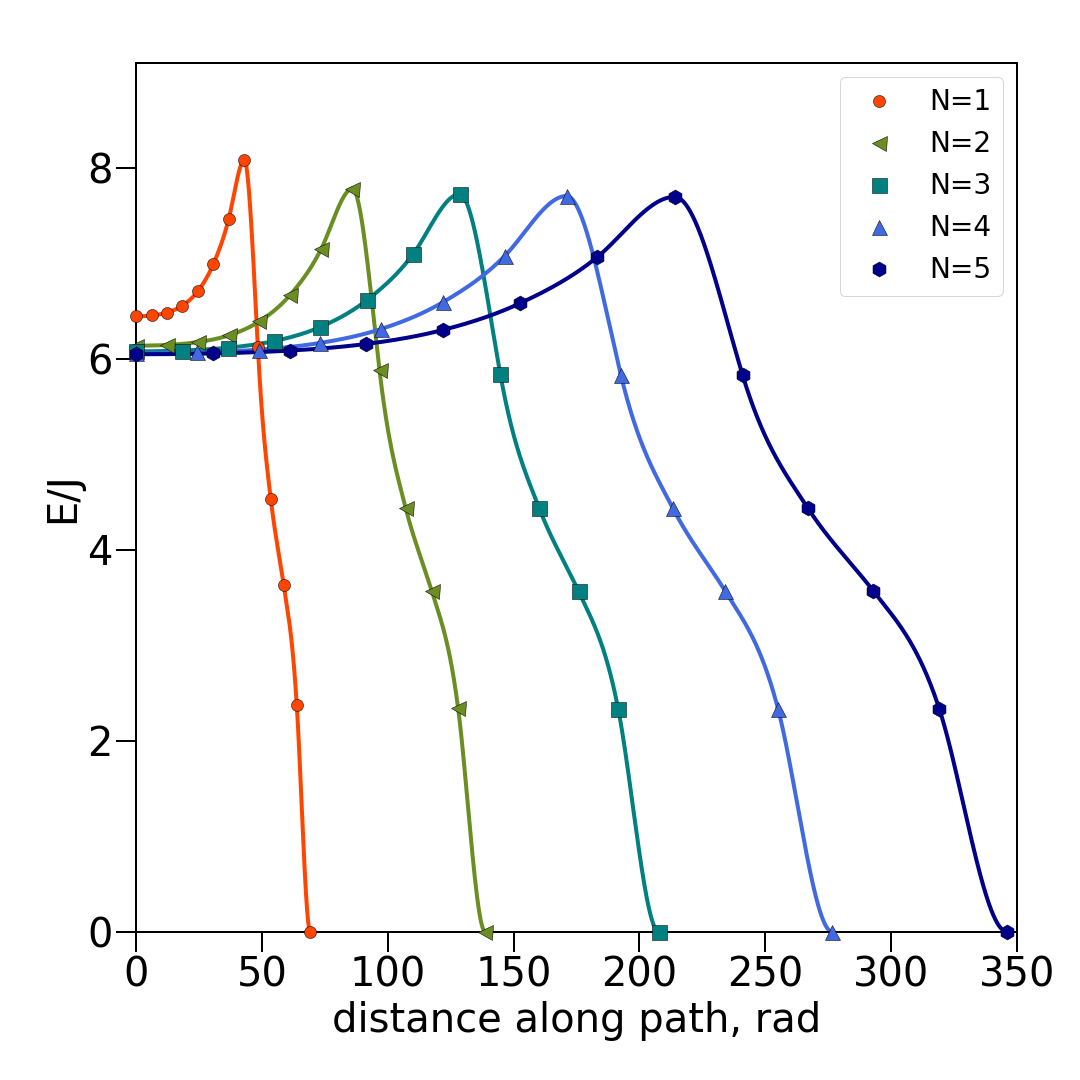}
\caption{Minimum energy path for skyrmion escape through the boundary of a track with finite width for several values of the scaling parameter, $N$. The activation energy reaches a limiting, constant value for small values of $N$. The zero of energy is taken to be that of the final ferromagnetic state.}
\label{MEPboundary}
\end{figure}
Figure \ref{MEPboundary} shows the MEP for skyrmion escape through the sample boundary for several values of the scaling parameter.
The width of the track for $N = 1$ corresponds to 30 atomic rows, about 3 times larger than the size of the skyrmion, wide enough for the boundaries not to have significant effect on the skyrmion state.
The energy barrier for escape first decreases slightly with $N$ 
but then remains constant. 
Thus, the activation energy for this skyrmion annihilation mechanism is nearly independent of scaling for $N>2$.
Unlike the collapse mechanism, escape can be described with a micromagnetic model \cite{Suess18}.


\subsection{Pre-exponential factor}

\begin{figure}[h]
\centering
\includegraphics[width=1\textwidth]{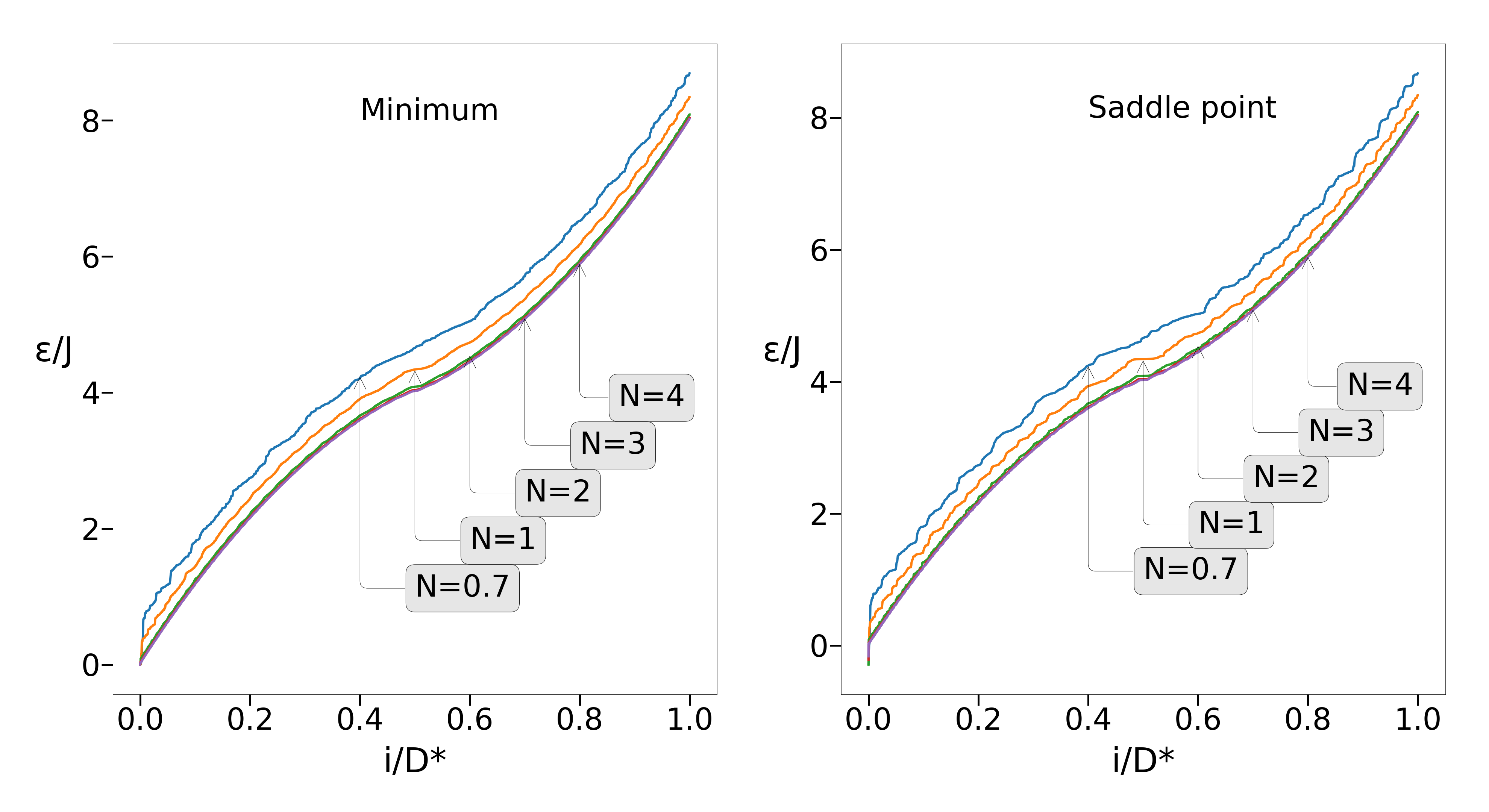}
\caption{Eigenvalues of the Hessian matrix for the initial skyrmion state (left) and at the saddle point corresponding to the transition state (right) as a function of $i/D^\ast$, where $i$ is a number of the eigenvalue when they are arranged in increasing order. $D^\ast$ is the total number of positive eigenvalues. The value of the scaling parameter $N$ for each curve is shown.}
\label{ei}
\end{figure}

Figure \ref{ei} shows calculated eigenvalues of the Hessian arranged in increasing order.
The horizontal axis gives the number of the eigenvalue in this sequence divided by the total number of eigenvalues, $i/D^\ast$. This is a monotone function, $\epsilon_i (i/D^\ast)$, on the interval $ [0,1]$, defined for a discrete set of points. 

As $N$ is increased, additional eigenvalues are introduced and the scaled index $i/D^\ast$ becomes more densely distributed over the interval $[0,1]$. As $N$ goes to infinity, the eigenvalues approach a limiting function $\epsilon (x)$. In figure \ref{ei} 
the left and right panels correspond to the initial skyrmion state and transition state respectively.  In both cases, convergence has been achieved for practical purposes already at $N=2$.

In the limit of large $N$, the expression in parentheses in the exponent in eq. (\ref{epsilon}) can be regarded as a Riemann sum for a definite integral over the function $\ln{\epsilon(x)}$ and, therefore,
\begin{equation}
\frac{1}{D^\ast}\sum\limits_{i}^{D^\ast} \frac{1}{2}\ln{\epsilon_i}\stackrel{{D^\ast} \to \infty}{\longrightarrow} \frac{1}{2} \int\limits_{0}^{1}\ln{\epsilon (x)}dx .
\label{limit}
\end{equation}
An expression of the form (\ref{limit}) enters both in the numerator (\ref{eq:kHTST2}) at the point corresponding to the skyrmion state, as well as into the denominator of eq. (\ref{eq:kHTST2}) at the saddle point.

If the integrals over $ \ln \epsilon(x)$ at the minimum and at the saddle point are different, an exponential dependence of the entropic contribution to $f_0$ on the scaling parameter $N$ is obtained. For relatively small value of $D^\ast$, the density of the eigenvalues of the Hessian has been analyzed previously \cite{Desplat18}, where the importance of accurate consideration of close to zero eigenvalues was emphasized. The case of large $D^\ast$ is computationally challenging since the calculation of the function $\epsilon(x)$ presupposes the knowledge of all eigenvalues of a matrix with a rank of several million with high accuracy already for $N \approx 100$.

The results presented here show that the entropy term in the pre-exponential factor for the lifetime, $\tau_0$, first 
increases as the scaling parameter $N$ increases, but then approaches a constant value for $ N> 30 $. 
The 
increase
in the entropy factor corresponds to five orders of magnitude while the dynamic term, containing the $b_j$ in eq. (\ref{eq:kHTST2}), changes by less than one order of magnitude.
A strong 
increase
in the pre-exponential factor with increasing skyrmion size explains the possible stability of micrometer skyrmions at room temperature even without taking into account the magnetic dipole interaction and can be considered as a manifestation of topological stabilization in the limit of an infinitesimal lattice constant.

The pre-exponential factor corresponding to skyrmion escape through the boundary also changes with the scaling parameter $ N $, but only by two orders of magnitude, and already for $ N> 4 $ it is practically constant. In contrast to the case of collapse in an extended system, the escape involves unequal number of zero modes at the minimum and at the saddle point (two for the skyrmion but only one for the transition state). The pre-exponential factor is, therefore, temperature dependent.
The variation of the lifetime due to escape with $N$ is much smaller than for skyrmion collapse inside the sample.


\section{Conclusions}
The activation energy and pre-exponential factor in the Arrhenius expression for the life time of a skyrmion is estimated within HTST for a set of decreasing values of the lattice constant while keeping the size of the skyrmion fixed. In the limit of infinitesimal lattice constant, the activation energy reaches a finite value, unlike the topological protection implied by a continuum model. The energy of the transition state approaches the Belavin-Polyakov limit of $4\pi J$ with respect to the ferromagnetic state.  An efficient method for calculating the pre-exponential factor for transitions in large systems is used where the eigenvalues of the Hessian need not be evaluated. The calculated results demonstrate that the entropy term in the pre-exponential factor decreases as the lattice constant is decreased but also approaches a constant value. The results can also be interpreted in terms of changing skyrmion size for a fixed lattice constant and then demonstrate how the decrease in the pre-exponential factor with increased skyrmion size makes an important contribution to the stability of micrometer skyrmions even in the absence of magnetic dipole interaction.


\vspace{0.25cm}
{\bf Acknowledgments}

\vspace{0.25cm}
This work was supported by Russian Science Foundation (Grant 19-42-06302), the Icelandic Research Fund, and the Research Fund of the University of Iceland.

 \bibliographystyle{elsarticle-num} 
 \bibliography{cas-refs}





\end{document}